\documentclass[11pt,a4paper]{article}
\pdfoutput=1
\usepackage{jheppub}
\usepackage[utf8]{inputenc} 
\usepackage{multirow}    
\usepackage{xcolor}
\usepackage{graphicx}
\usepackage{amsmath}
\usepackage{amssymb}
\usepackage{braket} 
\usepackage{hyperref}
\usepackage{frontespizio}
\usepackage{comment}  
\usepackage{epsfig}
\usepackage{float}
\usepackage{bbold}
\usepackage[title]{appendix}
\usepackage{multirow} 
\usepackage{tikz}
\usetikzlibrary{decorations.pathmorphing}
\usepackage{varwidth}
\usepackage{tikz}
\usetikzlibrary{backgrounds,patterns,calc}
\usepackage{xparse}  
\usepackage{subfigure}  
\usepackage{bbm}
\usepackage{wasysym}

\usepackage{jheppub} 

\title{Moore-Tachikawa Varieties: Beyond Duality }

\author[]{Veronica Pasquarella}
\affiliation{Department of Applied Mathematics and Theoretical Physics (DAMTP)}
\affiliation{University of Cambridge, Wilberforce Road, CB3 0WA, Cambridge, UK}

\emailAdd{vp360@damtp.cam.ac.uk}

\abstract{We propose a generalisation of the Moore-Tachikawa varieties for the case in which the target category of the 2D TFT is a hyperk$\ddot{\text{a}}$hler quotient. The setup requires generalising the bordism operators of Moore and Segal to the case involving lack of reparametrisation-invariance on the Riemann surface, ultimately enabling to relate this to the issue of defining a Drinfeld center for composite class ${\cal S}$ theories.}

\keywords{cochain level theories, symplectic varieties, topological field theories, categorical duality, quiver gauge theories, higher categories, relative QFTs}   
\makeatletter
\gdef\@fpheader{}
\makeatother

\begin{document} 
\maketitle

\section{Introduction}  

The present article is the first of a series by the same author meant to provide a mathematical explanation of the claims made in two previous works, \cite{Pasquarella:2023deo, Pasquarella:2023exd}. In doing so, we highlight interesting connections with theoretical physics results, quoting part of the most relevant literature wherever possible. 

Our main aim is understanding the underlining mathematical structure associated to the partition function, correlation functions, and spectrum of operators of a given quantum field theory (QFT). Specifically, we focus on supersymmetric gauge theories obtained by dimensional reduction of 6D ${\cal N}=(2,0)$ SCFTs, \cite{Witten:1995zh,Witten:2009at,DBZ,Witten:2007ct,Moore}. 

\begin{figure}[ht!]   
\begin{center}  
\includegraphics[scale=0.9]{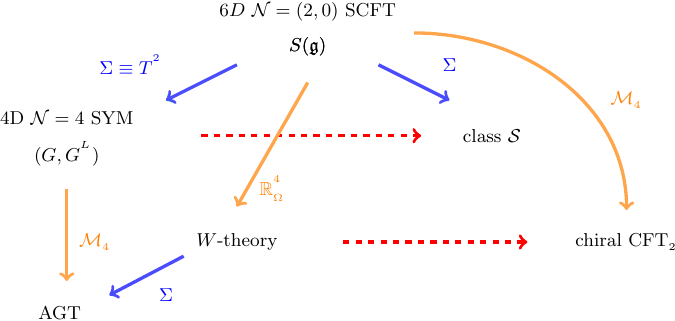}   
\caption{\small Partial reproduction of a diagram displayed in \cite{Moore}. The first part of our treatment focuses on the functorial field theory description of class ${\cal S}$ theories and their Higgs branches in terms of 2D TFT cobordism constructions.}    
\label{fig:classS}    
\end{center}  
\end{figure}

As explained in \cite{Pasquarella:2023deo, Pasquarella:2023exd}, for a theory, ${\cal T}$, to be absolute, the following triple needs to be defined

\begin{equation}   
\boxed{\ \ \ {\cal T}\ \longleftrightarrow\ ({\cal F}, \mu, \mathfrak{Z})\color{white}\bigg]\  },    
\label{eq:HB1}  
\end{equation}     
where ${\cal F}$ is the fiber functor, $\mu$ the moment map, and $\mathfrak{Z}$ the Drinfeld center of a given theory\footnote{We refer to \cite{Pasquarella:2023deo} for a detailed explanation of this terminology.}. Consistency of the underlying mathematical structure requires these three quantities to be mutually related. Indeed, upon defining any one of them, the other two should automatically follow. The purpose of \cite{Pasquarella:2023deo,Pasquarella:2023exd} and the present work is to show that an apparent shortcoming in defining such triple corresponds to the emergence of interesting physics, rather than being a fault of the sought after absolute theory. In particular, we will show that this can be used to explain the emergence of non-invertible symmetries separating different class ${\cal S}$ theories, \cite{Pasquarella:2023deo,Bashmakov:2022jtl,Bashmakov:2022uek,9,Kaidi:2021xfk,Choi:2022zal,Choi:2021kmx}. 


The crucial references we rely upon are the works of Moore and Segal, \cite{Moore:2006dw}, and Moore and Tachikawa, \cite{Moore:2011ee}. As briefly reviewed in the following sections, \cite{Moore:2011ee} proposes a redefinition of class ${\cal S}$ theories (cf. figure \ref{fig:classS}) in terms of a 2D TFT, namely the functor

\begin{equation}   
\boxed{\ \ \ \eta_{_{G_{\mathbb{C}}}}:\ \text{Bo}_{_2}\ \rightarrow\ \text{HS} \color{white}\bigg]\ \ \ }   
\label{eq:etaGC11}
\end{equation} 
with Bo$_{_2}$ and HS denoting the bordism 2-category and the holomorphic symplectic 2-category, respectively\footnote{For more details, we refer the reader to section \ref{sec:BMTCs}.}, associated to a given 4D ${\cal N}=2$ SCFT. The definition of \eqref{eq:etaGC11} strongly relies upon assuming, both, the source and target categories, enjoy a duality structure which, in turn follows from the presence of an identity element in both categories. In \cite{Moore:2011ee}, the authors show that, under the duality assumption, for the categories in \eqref{eq:etaGC11} to be well-defined, it is enough to specify their objects and 1-morphisms. Essentially, the objects of Bo$_{_2}$ are circles, $S^{^1}$, and the 1-morphisms are cobordisms between different disjoint unions of circles and the empty set. Their respective counterpart on the holomorphic symplectic side correspond to the gauge group, \cite{Moore:2011ee},

\begin{equation}    
\eta_{_{G_{_{\mathbb{C}}}}}\ \left(S^{^1}\right)\ \overset{def.}{=}\ G_{_{\mathbb{C}}},  
\label{eq:gaugegroup}   
\end{equation}  
and the cobordism operators, \cite{Moore:2011ee},
\begin{equation}    
\eta_{_{G_{_{\mathbb{C}}}}}\ \left(\text{Hom}\ \left(S^{^1}, \emptyset\right)\right)\ \overset{def.}{=}\ U_{_{G_{_{\mathbb{C}}}}}
\label{eq:UG1} 
\end{equation}  
\begin{equation}
\eta_{_{G_{_{\mathbb{C}}}}}\ \left(\text{Hom}\ \left(S^{^1}\ \sqcup\ S^{^1}, \emptyset\right)\right)\ \overset{def.}{=}\ V_{_{G_{_{\mathbb{C}}}}}  
\label{eq:VG1} 
\end{equation}  

\begin{equation} 
\eta_{_{G_{_{\mathbb{C}}}}}\ \left(\text{Hom}\ \left(S^{^1}\ \sqcup\ S^{^1}\ \sqcup \ S^{^1},\emptyset\right)\right)\ \overset{def.}{=}\ W_{_{G_{_{\mathbb{C}}}}}.    
\label{eq:HBCS}   
\end{equation}  
respectively.

Importantly for us, the Moore-Tachikawa varieties described by \eqref{eq:gaugegroup}, \eqref{eq:UG1}, \eqref{eq:VG1}, and \eqref{eq:HBCS} constitute the quiver gauge theory\footnote{Among the cobordism operators is the Higgs branch of class ${\cal S}$ theories, \eqref{eq:HBCS}.} realisation of \cite{Moore:2006dw}, where Moore and Segal propose the mathematical formalism needed for addressing the following question: given a certain closed string theory background, what is its corresponding D-brane content\footnote{Note that this is essentially the same question addressed in \cite{Aharony:2013hda}.}?

Our aim is that of explaining how and why one needs to generalise the construction of \cite{Moore:2011ee} from a higher-categorical point of view, and what its implications are on the theoretical physics side. In doing so, we highlight the crucial properties and axioms satisfied by the cobordism operators outlined in \cite{Moore:2006dw,Moore:2011ee}, and how they can be generalised to account for more general setups from, both, the mathematical and theoretical physics perspectives.

In particular, we emphasise the dependence of \eqref{eq:etaGC11} on the conformal structure of the Riemann surface on which the compactification of the 6D ${\cal N}=(2,0)$ SCFT has been performed to achieve a certain class ${\cal S}$ theory and how lack of reparametrisation invariance, corresponding to the absence of the identity element in its source and target categories, \cite{Moore:2011ee}, signals the presence of (non-invertible) categorical symmetries separating different absolute theories. 

\begin{figure}[ht!]   
\begin{center}  
\includegraphics[scale=0.75]{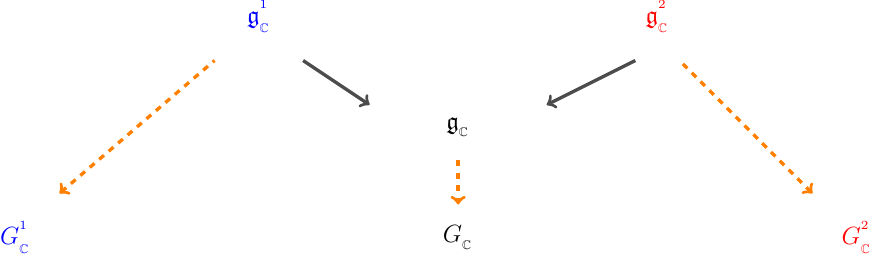}   
\caption{\small Adaptation of a correspondence first proposed in \cite{Pasquarella:2023deo} playing a key role towards generalising \cite{Moore:2011ee} to the hyperk$\ddot{\text{a}}$hler target category case. As explained in section \ref{sec:new1234}, this also requires the generalisation of cobordism operators, \cite{Moore:2006dw}, due to the lack of reparametrisation-invariance of the Riemann surface on which the compactification of the 6D ${\cal N}=(2,0)$ SCFT is performed.}    
\label{fig:correspondence}    
\end{center}  
\end{figure}

At the heart of this is the correspondence sketched in figure \ref{fig:correspondence}.

In \cite{Pasquarella:2023deo} we explained how gauging a Symmetry Topological Field Theory (SymTFT) enables to change the boundary conditions of the fields living in the absolute field theory resulting from the Freed-Moore-Teleman construction, \cite{Freed:2012bs,Freed:2022qnc,Freed:2022iao,TJF,Kong:2013aya,Yu:2021zmu}. To each gauging corresponds a choice of triples, \eqref{eq:etaGC11}, and, for any absolute theory, it is enough to define one of the three entries on the RHS of \eqref{eq:etaGC11} to determine the other two. For the purpose of this article, we will mostly focus on the second, namely the moment  map, defined as follows

\begin{equation} 
\mu:\ \mathfrak{G}\ \rightarrow\ {\cal A},      
\end{equation} 
where $\mathfrak{G}$ is an n-categorical structure, and ${\cal A}$ is the algebra of invertible topological defects associated to the action $\mu\ (\mathfrak{G})$. Gauging means taking the categorical quotient with respect to $\mathfrak{G}$ (in notation $//_{_{\mu}}\ \mathfrak{G}$), and projecting its image under $\mu$ to the identity\footnote{We thank Nathan Seiberg for instructive discussion regarding the appropriateness of the terminology to be adopted in describing this formalism.}. Practically, one could perform a total gauging of the theory by choosing ${\cal A}$ such that the overall spectrum of the theory of the gauged theory is only the (new) identity element. For the purpose of our work, instead, we are interested in understanding mathematical structures arising by gauging with respect to different subalgebras within ${\cal A}$, that are mutually intersecting, albeit not contained within each other. The ultimate aim is that of explaining the emergence of (non-invertible) categorical symmetries in certain supersymmetric quiver gauge theories once described in terms of Coulomb branches of magnetic quivers of 3D ${\cal N}=4$ quiver gauge theories, which is the main focus of an upcoming paper by the same author, \cite{VP}. In such analysis we will be applying some of the findings of \cite{Braverman:2017ofm,Benini:2010uu}.

This article is structured as follows: section \ref{sec:Moore-Segal} is devoted to a brief overview of cochain level theories as the most important generalisation of the open and closed TFT construction. Mostly relying upon \cite{Moore:2006dw}, we highlight the importance of the construction of the cobordism operator, highlighting its dependence on the conformal structure of the Riemann surface. In section \ref{sec:BMTCs} we then turn to the discussion of a particular 2D TFT valued in a symmetric monoidal category, namely the maximal dimension Higgs branch of class ${\cal S}$ theories. After briefly reviewing the properties outlined in \cite{Moore:2011ee}, in section \ref{sec:new1234} we propose their generalisation for the case in which the target category of the $\eta_{_{G_{_{\mathbb{C}}}}}$ functor is a hyperk$\ddot{\text{a}}$hler quotient. We conclude outlining the possible extension of this treatment towards a mathematical formulation of magnetic quivers within the context of Coulomb branches of 3D ${\cal N}=4$ quiver gauge theories which will be addressed in an upcoming work by the same author, \cite{VP}.

\section{Cochain level theories}   \label{sec:Moore-Segal}  

This first section is devoted to a brief overview of cochain level theories as the most important generalisation of the open and closed TFT construction. The reason for doing so is that these mathematical structures are central to the idea of D-branes, \cite{Moore:2006dw}, enabling to determine the set of possible D-branes given a closed string background. In their work, \cite{Moore:2006dw} address this problem from the point of view of a 2D TFT, \eqref{eq:etaGC11}, where the whole content of the theory is encoded in a finite-dimensional commutative Frobenius algebra. 






The present section is therefore structured as follows:   

\begin{enumerate}  

\item At first, we briefly overview cochain complexes as the essential mathematical tools needed for translating the setup of our previous work, \cite{Pasquarella:2023deo}, in the formalism of Moore and Segal. 

\item We then turn to highlighting the construction of cobordism operators, \cite{Moore:2006dw}, emphasising its dependence on the conformal structure of the Riemann surface.

\end{enumerate}

\subsection{Cochain complexes}

A cochain complex, $(A^{^{\tiny\bullet}}, d^{^{\tiny\bullet}})$, is  an algebraic structure that consists of a sequence of abelian groups (or modules), $A^{^{\tiny\bullet}}$, and a sequence of homomorphisms between consecutive groups, $d^{^{\tiny\bullet}}$, such that the image of each homomorphism is included in the kernel of the next. To a chain complex, $(A_{_{\tiny\bullet}}, d_{_{\tiny\bullet}})$, there is an associated homology, which describes how the images are included in the kernels. A cochain complex is similar to a chain complex, except that its homomorphisms are in the opposite direction. The homology of a cochain complex is called its cohomology

\begin{equation}  
H({\cal C})\ \overset{def.}{=}\ \text{Ker}({\cal Q})/ \ \text{Im}({\cal Q}).
\end{equation}  

The n$^{th}$ cohomology group, $H_{_n}(H^{^0})$ is 

\begin{equation}  
H_{_n}\ \overset{def.}{=}\ \text{Ker}\ d_{_n}/ \ \text{Im}\ d_{_{n+1}}.
\end{equation} 

The central object in closed string theory is the vector space ${\cal C}\equiv{\cal C}_{_{S^{^1}}}$ of states of a single parametrised string. ${\cal C}$ denotes the cochain complex in this case, \cite{Moore:2006dw}. The latter comes equipped with a grading given by the ghost number, and an operator ${\cal Q}:\ {\cal C}\ \rightarrow\ {\cal C}$ called the BRST operator, raising the ghost number by 1, and such that ${\cal Q}^{^2}\equiv 0$.

\subsection{The Moore-Segal setup}   

The most general finite-dimensional commutative algebra over the complex numbers is of the form

\begin{equation}  
{\cal C}\ \overset{def.}{=}\ \underset{x}{\bigoplus}\ {\cal C}_{_{x}}  \ \ \ , \ \ \ x\in\ \text{Spec}\ ({\cal C}),  
\end{equation}
with 

\begin{equation}  
{\cal C}_{_{x}}\ \overset{def.}{=}\ \mathbb{C}\ {\cal E}_{_{x}}\ \oplus\ m_{_{x}},    
\end{equation}  
where ${\cal E}_{_{x}}$ is an idempotent, and $m_{_{x}}$ a nilpotent ideal. If ${\cal C}$ is a Frobenius algebra, then so too is each ${\cal C}_{_{x}}$.

In their treatment, \cite{Moore:2006dw} restrict to the semisimple\footnote{Despite appearing quite restrictive, committing to semisimplicity is enough to shed light on the essential structure of the theory. According to \cite{Moore:2006dw}, to go beyond it, the appropriate objects of study, are cochain-complex valued TFTs rather than non-semisimple TFTs in the usual sense.}  case. Semisimplicity admits many equivalent definitions:

\begin{enumerate}   

\item  The presence of simultaneously-diagonalisable fusion rules.   

\item  There exists a set of basic idempotents ${\cal E}_{_{x}}$ such that 
\begin{equation}  
{\cal C}\ \overset{def.}{=}\ \underset{x}{\bigoplus}\ \mathbb{C}\ {\cal E}_{_{x}}  \ \ \ , \ \ \ x\in\ \text{Spec}\ ({\cal C}),\ \ \ \text{with}\ \ \ \ \   
{\cal E}_{_{x}}{\cal E}_{_{y}}\ \equiv\ \delta_{_{xy}} {\cal E}_{_{y}}.   
\end{equation}

\item   ${\cal C}$ is the algebra of complex-valued functions on the finite set of characters of ${\cal C}$, $X\in\ \text{Spec}\ ({\cal C})$.   

\end{enumerate}

For any pair of boundary conditions, $a,b$, the corresponding cochain complex for a semisimple category is defined as follows, \cite{Moore:2006dw},

\begin{equation}  
{\cal O}_{_{aa}}\ \simeq\ \underset{x}{\bigoplus}\ \text{End}\left(W_{_{x,a}}\right),     
\end{equation}

\begin{equation}  
{\cal O}_{_{ab}}\ \simeq\ \underset{x}{\bigoplus}\ \text{Hom}\left(W_{_{x,a}}; W_{_{x,b}}\right),      
\end{equation}   
where $W_{_{x,a}}$ is a vector space associated to every idempotent ${\cal E}_{_{x}}$.

\subsection{Cobordism operators}   

We now turn to the key elements for our analysis. In this first section we will be using the definition provided by \cite{Moore:2006dw}, highlighting the crucial property that will be mostly needed in sections \ref{sec:BMTCs} and \ref{sec:new1234}. A cobordism $\Sigma$ from $p$ circles to $q$ circles gives an operator   

\begin{equation}   
U_{_{\Sigma, \alpha}}:\ {\cal C}^{^{\otimes p}}\ \rightarrow\ {\cal C}^{^{\otimes q}},  
\label{eq:U}  
\end{equation}   
which depends on the conformal structure $\alpha$ on $\Sigma$. This operator, \eqref{eq:U}, is a cochain map, but its crucial feature is that, changing the conformal structure $\alpha$ on $\Sigma$, changes $U_{_{\Sigma, \alpha}}$ only by a cochain homotopy.   

To describe how $U_{_{\Sigma, \alpha}}$ varies with $\alpha$, if ${\cal M}_{_{\Sigma}}$ is the moduli space of conformal structures on the cobordism $\Sigma$ which are the identity on the boundary circles, there is a resulting cochain map    

\begin{equation}  
U_{_{\Sigma}}:\ {\cal C}^{^{\otimes p}}\ \rightarrow\ \Omega\left({\cal M}_{_{\Sigma}}; {\cal C}^{^{\otimes q}}\right),     
\end{equation}  
with the target denoting the de Rham complex of forms on ${\cal M}_{_{\Sigma}}$ with values in ${\cal C}^{^{\otimes q}}$. 

An alternative equivalent definition is that of the following cochain map

\begin{equation}  
U_{_{\Sigma}}:\ C_{_{_{_{\tiny\bullet}}}}\ \left( {\cal M}_{_{\Sigma}}\right)\ \rightarrow\ \left({\cal C}^{^{\otimes p}}\right)^{^*}\ \otimes\ {\cal C}^{^{\otimes q}}.   
\end{equation} 

In the next sections, we will show that lack of reparametrisation-invariance of the Riemann surface implies interesting mathematical and physical features of the resulting theory of interest.

\section*{Key points}    

The main points to keep in mind throughout the reminder of ourtreatment are the following:  

\begin{itemize}  

\item Cochain level theories provide the natural mathematical formalism for describing absolute theories obtained by partial gaugings of the SymTFT in the Freed-Moore-Teleman setup. 

\item The definition of cobordism operators associated to such complex cochain structure follows from the assumption that the Riemann surface is reparametrisation-invariant. 

\end{itemize}

\section{Moore-Tachikawa varieties}   \label{sec:BMTCs}

Having outlined the importance of reparametrisation-invariance in the definition of bordism operators, \cite{Moore:2006dw}, we now turn to the particular application in describing maximal dimensional Higgs branches of class ${\cal S}$ theories, as first proposed by \cite{Moore:2011ee}. Our major contribution in the present section will be highlighting where upgrades to the categories defined in \cite{Moore:2006dw} are need for dealing with setups as the ones associated to the correspondence depicted in figure \ref{fig:correspondence}, namely those leading to the emergence of composite class ${\cal S}$ theories separated by a non-invertible defect. 

This section is structured as follows:   

\begin{enumerate}

\item  At first, we briefly overview the source and target categorical structure proposed in \cite{Moore:2011ee} assuming duality.   

\item  We then explain what categorical duality means from an algebraic perspective.

\item  We conclude the section indicating the relation between Moore-Tachikawa varieties and Coulomb branches of quiver gauge theories as an interesting realisation of 3D mirror symmetry, and how the categorical generalisation proposed in this work suggests interesting applications to quiver varieties that will be addressed in more detail in \cite{VP}.

\end{enumerate}

\subsection{Categorical structure assuming duality}  

As already mentioned in the Introduction, to a given class ${\cal S}$ theory, one can assign a 2D TFT valued in a symmetric monoidal category, \cite{Moore:2011ee},

\begin{equation}   
\boxed{\ \ \ \eta_{_{G_{\mathbb{C}}}}:\ \text{Bo}_{_2}\ \rightarrow\ \text{HS} \color{white}\bigg]\ \ \ }   
\label{eq:etaGC}
\end{equation}

The existence of this 2D TFT relies on the the source and target categories satisfying a certain list of properties, \cite{Moore:2011ee}. We will not reproduce all of them in our treatment, and refer the interested reader to the original work of Moore and Tachikawa for a detailed explanation. In this first part of the section, we will only point out some of the crucial assumptions made in their work for reasons that will become clear in the following pages. 

\section*{Duality}

\begin{figure}[ht!]       
\begin{center}  
\includegraphics[scale=1]{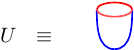}   \ \ \  \ \ \ 
  \includegraphics[scale=1]{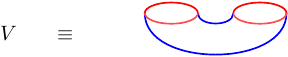} \ \ \   \ \ \ 
\includegraphics[scale=1]{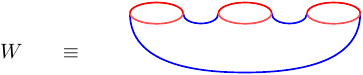}   
\caption{\small Basic bordisms assuming duality of, both, the source and target categories leading to the definition of the identity element, $V_{_{G_{_{\mathbb{C}}}}}$, and the maximal dimensional Higgs branch, $W_{_{G_{_{\mathbb{C}}}}}$.}    
\label{figure:UW}    
\end{center}  
\end{figure}

For the purpose of our work, the crucial assumption made in \cite{Moore:2011ee} is the duality structure of the source category Bo$_{_2}$. As explained in \cite{Moore:2011ee}, duality implies that the 2-category Bo$_{_2}$ is fully specified by its objects, $S^{1}$, and 1-morphisms, namely the bordisms depicted in figure \ref{figure:UW}. The middle bordism, i.e. the one labelled $V$, is the identity bordism. One can easily see this by noticing that $V$ is topologically equivalent to a cylinder whose edges are the red circles, i.e. the object of 2-category Bo$_{_2}$ (the closed string we were referring to in section \ref{sec:Moore-Segal}).

For $\eta_{_{G_{\mathbb{C}}}}$ to be well defined, the source and target categories are required to satisfy certain sewing relations, \cite{Moore:2006dw,Moore:2011ee}. This practically means that, compositions between morphisms should close. In particular, the identity itself can be defined in terms of composite homomorphisms as follows,

\begin{equation}             
\boxed{\ \ \ U_{_{G_{\mathbb{C}}}}\ \circ\ W_{_{G_{\mathbb{C}}}}\ \equiv\ T^{^*}G_{_{\mathbb{C}}} \color{white}\bigg]\ },  
\label{eq:defofid}     
\end{equation} 
where $T^{^*}G_{_{\mathbb{C}}}\ \equiv\ V{_{G_{\mathbb{C}}}}$.
Indeed, one can easily see that combining the first and third bordims in figure \ref{figure:UW}, is topologically equivalent to $V$.

\footnote{Indeed, $V$ is topologically equivalent to the cylinder, i.e. the cobordism between $S^{^1}$ and itself.}

We therefore wish to highlight the following  

 \medskip    
   \medskip
\color{blue}

\noindent\fbox{%
    \parbox{\textwidth}{%
   \medskip    
   \medskip
   \begin{minipage}{20pt}
        \ \ \ \ 
        \end{minipage}
        \begin{minipage}{380pt}
      \color{black}  \underline{Main point:} Duality ensures the presence of an identity associated to a certain gauge group, $G_{_{\mathbb{C}}}\equiv\eta_{_{G_{\mathbb{C}}}}\ (S^{^1})$, \eqref{eq:gaugegroup}.
        \end{minipage}   
         \medskip    
   \medskip
        \\
    }%
}
 \medskip    
   \medskip     \color{black}

\subsection*{Key axiom}  

\eqref{eq:defofid} is essential for us in relating the formalism of \cite{Moore:2006dw,Moore:2011ee} to the setup of figure \ref{fig:correspondence}. In particular, it is what leads to the definition of the triple featuring on the RHS of \eqref{eq:HB1}. To see this explicitly, let us recall a crucial axiom required to be satisfied by \eqref{eq:etaGC}, and, therefore, in turn by \eqref{eq:defofid}, \cite{Moore:2011ee}.

For $X\in$\ Hom\ $(G_{_{\mathbb{C}}}^{^{\prime}}, G_{_{\mathbb{C}}})$ and $Y\in$\ Hom\ $(G_{_{\mathbb{C}}}, G_{_{\mathbb{C}}}^{^{\prime\prime}})$, their composition 

\begin{equation}   
Y\ \circ\ X\ \in\ \text{Hom}\ (G_{_{\mathbb{C}}}^{^{\prime}}, G_{_{\mathbb{C}}}^{^{\prime\prime}}) 
\label{eq:etaGC1}
\end{equation}    
is identified with the holomorphic symplectic quotient   

\begin{equation}   
\begin{aligned}
Y\ \circ\ X\ &\overset{def.}{=}\ X\ \times\ Y\ //\ G_{_{\mathbb{C}}}   \\
&\ =\{(x,y)\ \in\ X\times Y|\ \mu_{_X}(x)+\mu_{_Y}(y)=0\}\ /\ G_{_{\mathbb{C}}},     
\end{aligned}
\label{eq:etaGC1}
\end{equation}  
where

\begin{equation}   
\mu_{_X}:\ X\ \longrightarrow\ \mathfrak{g}_{_{\mathbb{C}}}^{*}\ \ \ , \ \ \ \mu_{_Y}:\ Y\ \longrightarrow\ \mathfrak{g}_{_{\mathbb{C}}}^{*}
\label{eq:etaGC1b}
\end{equation}  
are the moment maps of the action of $G_{_{\mathbb{C}}}$ on $X$ and $Y$, with $\mathfrak{g}_{_{\mathbb{C}}}$ the Lie algebra associated to $G_{_{\mathbb{C}}}$.  The identity element

\begin{equation}   
T^{^*}G_{_{\mathbb{C}}}\ \overset{def.}{=}\ \text{id}_{_{G_{_{\mathbb{C}}}}}\ \in\ \text{Hom}\ (G_{_{\mathbb{C}}}, G_{_{\mathbb{C}}}) 
\label{eq:etaGC2}
\end{equation}           
comes with a Hamiltonian $G_{_{\mathbb{C}}}\ \times\ G_{_{\mathbb{C}}}$ action. As also explained in \cite{Moore:2011ee}, to see that $T^{^*}G_{_{\mathbb{C}}}$ acts as the identity, it is enough to consider a composition of homomorphisms, $T^{^*}G_{_{\mathbb{C}}}\ \circ\ X$. Identifying $T^{^*}G_{_{\mathbb{C}}}\ \simeq\ G_{_{\mathbb{C}}}\ \times\ \mathfrak{g}_{_{\mathbb{C}}}$, and identifying an element of $T^{^*}G_{_{\mathbb{C}}}$ as $(g,a)$.   The moment map condition, \eqref{eq:etaGC1}, reduces to   

\begin{equation}   
a+\mu(x)=0,      
\label{eq:etaGC3}
\end{equation}  
from which $a$ can be removed. Consequently, the induced 2-form on the solution space is $G_{_{\mathbb{C}}}$-invariant and basic. Upon taking the quotient with respect to $G_{_{\mathbb{C}}}$, we can gauge $g$ to 1, leading to a holomorphic isomorphism with the original $X$ space with its symplectic form.

The categorical quotient taken in defining the composition \eqref{eq:etaGC1} is equivalent to the one that a given absolute theory should be equipped with to potentially gauge away its entire operator content, while leaving only the identity in the spectrum\footnote{cf. explanation in the Introduction. This is basically what leads to the definition of the fiber functor, moment map and Drinfeld center.}. Indeed, this is true as long as the embeddings of the subalgebras associated to $G_{_{\mathbb{C}}}^{^{\prime}}$ and $G_{_{\mathbb{C}}}^{^{\prime\prime}}$ are subsets of each other within the mother algebra $\mathfrak{g}_{_{\mathbb{C}}}$. However, we are interested in describing more general setups, where the embeddings of the algebras are intersecting albeit not one included within the other. In the remainder of our treatment, we will explain that, for this to be described in the formalism of \cite{Moore:2006dw,Moore:2011ee}, the standard identity element associated to the gauge group $G_{_{\mathbb{C}}}$ and embedding Lie algebra $g_{_{\mathbb{C}}}$ needs to be removed from Bo$_{_2}$, while being replaced by a new composite bordism, and propose the definition of a new functor.

\subsection{Duality from an algebraic perspective} \label{sec:od}      

Before turning to explaining what are the changes that the soruce and target categories should undergo\footnote{Which will be the core topic of section \ref{sec:new1234}.}, we will briefly pause for a digression explaining how reparametrisation-invariance of the Riemann surface involved in the definition of the bordism operators, outlined in section \ref{sec:Moore-Segal}, is strongly related to the aforementioned duality assumption.

As explained in \cite{Moore:2011ee}, the crucial point is that, thanks to the duality propriety of the source 2-category Bo$_{_2}$, the identity element in the target category, $T^{^*}G_{_{\mathbf{C}}}$, is reparametrisation-invariant.  In particular, one could compose the identity morphisms as follows\footnote{Making use of the axiom \eqref{eq:etaGC1}.}  

\begin{equation}  
T^{^*}G_{_{\mathbf{C}}}^{^a{^{^\prime}}}\ \circ\ T^{^*}G_{_{\mathbf{C}}}^{^a}\equiv\left(T^{^*}G_{_{\mathbf{C}}}^{^a}\ \times\ T^{^*}G_{_{\mathbf{C}}}^{^a{^{^\prime}}} \right) //  G_{_{\mathbf{C}}}.    
\label{eq:composition}  
\end{equation}

From the considerations made above, it therefore follows that one could rephrase \eqref{eq:composition} as the definition of the Drinfeld center for the composite system made up of two class $\cal S$ theories (associated to the two gauge groups involved) separated by an invertible defect, with the latter ensuring reparametrisation invariance of the Riemann surface \cite{Moore:2006dw}.   

Concretely, under the duality assumption, one could gauge away either of the two groups, while being left with the following

\begin{equation}  
\mathbb{1}_{_{G_{_{\mathbf{C}}}^{^{a^{^\prime}}}}}\ \circ\ T^{^*}G_{_{\mathbf{C}}}^{^a}\equiv\left(T^{^*}G_{_{\mathbf{C}}}^{^a}\ \times\ \mathbb{1}_{_{G_{_{\mathbf{C}}}^{^{a^{^\prime}}}}} \right)//G_{_{\mathbf{C}}}.      
\end{equation}

If the conformal structure were the same on the two sides, then it would be the same, either with or without the composition rule. If 

\begin{equation}  
G_{_{\mathbf{C}}}^{^a}\ \times \ G_{_{\mathbf{C}}}^{^{a^{^\prime}}}\ \equiv\ G_{_{\mathbf{C}}}^{^{a+a^{^\prime}}}\ \equiv\ G_{_{\mathbf{C}}}, \ \ \ \forall a, a^{\prime},    
\end{equation}   
then \eqref{eq:composition} can be recast to the following

\begin{equation}  
\boxed{\ \ \ T^{^*}G_{_{\mathbf{C}}}^{^a}\equiv T^{^*}G_{_{\mathbf{C}}}^{^a}\ \  //  \ \ G_{_{\mathbf{C}}} \color{white}\bigg]\ \ } \ ,      
\label{eq:composition1}  
\end{equation}  
which is equivalent to a statement of S-duality. Once more, we highlight that this is possible because the source category for the 2D TFT associated to the group $G_{_{\mathbf{C}}}$ contains the identity element. But, in case this is not true\footnote{Such as the case in which the Riemann surface is no longer reparametrisation-invariant.}, \eqref{eq:composition1} needs to be changed accordingly, which one could think of as a generalisation of an S-duality statement. Indeed, if the group composition rules don't hold, 

\begin{equation}  
G_{_{\mathbf{C}}}^{^{a+a^{^\prime}}}\ \neq\ G_{_{\mathbf{C}}}, \ \ \ \forall a, a^{\prime},  
\end{equation}
we get something that is not simply the ordinary S-dual theory, \eqref{eq:composition1}.

The main purpose of our work is basically to go backwards, starting from the LHS of \eqref{eq:composition} and determining what the RHS should be. Most importantly, we need to:   

\begin{enumerate}  

\item Identify $G_{_{\mathbf{C}}}$ in the new theory obtained by composing the two theories on the LHS, each one characterised by a different choice of conformal structure on the Riemann surface.  

\item  Equivalently to 1., reconstruct $T^{^*}G_{_{\mathbf{C}}}$, namely the identity of the composite theory.

\item  We highlight that the most important generalisation of the 2D TFT \eqref{eq:etaGC1} that one should really be using for the case of interest to us is instead the following

\begin{equation}   
\boxed{\ \ \ \tilde\eta_{_{G_{\mathbb{C}}}}:\ \text{Bo}_{_2}\backslash\ V \rightarrow\ \text{HK} \color{white}\bigg]\ \ \ } 
\label{eq:newetaGC}   
\end{equation}    
which, as already pointed out in \cite{Moore:2011ee}, requires removing the identity element from the source category. Its effect on the target is to turn it into a hyperk$\ddot{\text{a}}$hler quotient. Its connection with theoretical physics\footnote{Already presented in \cite{Pasquarella:2023exd}.} is the main focus of \cite{VP}.   

\end{enumerate} 

From step 1., an important observation is in order. $G_{_{\mathbf{C}}}$ acts on the two factors on the RHS of \eqref{eq:composition} in separate ways. This is part of the meaning of the generalisation of S-duality that we were previously referring to. Indeed, the categorical quotient $//G_{_{\mathbf{C}}}$ tells us what the identity is as a result of gauging a certain subalgebra. This is obtained by taking the 1-morphisms on either side of the correspondence and taking their nontrivial composition w.r.t. $T^{^*}G_{_{\mathbf{C}}}$, with the latter being the identity in the target category. But the latter was assumed to be removed. The immediate suggestion to circumvent this shortcoming is that the functors defining the identity element of each individual theory on the LHS of \eqref{eq:composition} is different w.r.t. the one on the RHS. In section \ref{sec:new1234} we therefore propose the generalisation of S-duality as the need to define two different 2D TFT functors associated to the left and right hand sides of \eqref{eq:composition}.

\subsection{Algebraic Varieties} \label{sec:FBFN}

In the concluding part of this section, we highlight an interesting application of \eqref{eq:newetaGC} in the context of quiver gauge theories, which will be explained in more detail in an upcoming work by the same author, \cite{VP}, in particular towards generalising 3D mirror symmetry.

The Higgs branches described by Moore and Tachikawa are known to have been reproduced by \cite{Braverman:2017ofm} as the Coulomb branches of 3D ${\cal N}=4$ supersymmetric quiver gauge theories. Such correspondence is therefore equivalent to a statement of 3D mirror symmetry. The purpose of \cite{VP} is to explain what the 3D dual of a theory described by 
\eqref{eq:newetaGC} is in terms of Coulomb branches of 3D ${\cal N}=4$ quiver gauge theories. In this way, we expect to be able to prove the statements made in \cite{Pasquarella:2023exd}.

If $V$ has been removed from the source, one should expect there to be more than one 2D TFT 
 of the kind \eqref{eq:etaGC} associated to two different gauge groups whose embedding in the gauge group associated to the original TFT with identity element $V$ is not simply a cochain complex. Correspondingly, this also means that there is more than one 1-morphism $U_{_{G_{\mathbb{C}}}}$.

Given that the identity of the embedding theory is defined as follows

\begin{equation}             
\boxed{\ \ \ U_{_{G_{\mathbb{C}}}}\ \circ\ W_{_{G_{\mathbb{C}}}}\ \equiv\ T^{^*}G_{_{\mathbb{C}}} \color{white}\bigg]\ }\ \ \ \ , \ \ \ \  
\boxed{\ \ \ \eta_{_{G_{\mathbb{C}}}}(V)\ \equiv\ T^{^*}G_{_{\mathbb{C}}}\color{white}\bigg]\ \ }.    
\end{equation} 
and that 

\begin{equation}             
U_{_{G_{\mathbb{C}}}}\ \overset{def.}{=}\ G_{_{\mathbb{C}}}\times S_{_{n}}\ \subset\ G_{_{\mathbb{C}}}\ \times\ \mathfrak{g}_{_{\mathbb{C}}}\ \simeq\ T^{^*}G_{_{\mathbb{C}}}.    
\end{equation} 
with $S_{_n}$ is the Slodowy slice at a principal nilpotent element $n$. The physical theories of class ${\cal S}$ predict the existence of a variety $W_{_{G_{_{\mathbb{C}}}}}$ satisfying the properties needed to define a TFT $\eta_{_{G_{\mathbb{C}}}}$.  From the duality assumption, it follows that the dimensionalities of the two varieties are related as follows

\begin{equation}             
\text{dim}_{_{\mathbb{C}}}\ U_{_{G_{\mathbb{C}}}}\ \overset{def.}{=}\ \text{dim}_{_{\mathbb{C}}}\ G_{_{\mathbb{C}}}\ +\ \text{rank}\ G_{_{\mathbb{C}}}.  
\end{equation} 

\begin{equation}             
\text{dim}_{_{\mathbb{C}}}\ W_{_{G_{\mathbb{C}}}}\ \overset{def.}{=}\ 3\ \text{dim}_{_{\mathbb{C}}}\ G_{_{\mathbb{C}}}\ -\ \text{rank}\ G_{_{\mathbb{C}}}. 
\label{eq:dimhb}   
\end{equation} 

However, if the identity needs to be removed from Bo$_{_2}$, $T^{^*}G_{_{\mathbb{C}}}$ is not the identity and, in particular \eqref{eq:dimhb} needs to be redefined precisely because the source is no longer a dual category. How to rederine \eqref{eq:dimhb} will be explained in \cite{VP}.

As a concluding remark to what we have just said, in \cite{Moore:2006dw} they conjecture the following property for the moment maps associated to the $G^{^3}$ action on the Higgs branch $W_{_{G_{_{\mathbb{C}}}}}$

\begin{equation}   
\boxed{\ \ \ \mu_{_i}:\ W_{_{G_{_{\mathbb{C}}}}}\rightarrow \mathfrak{g}_{_{\mathbf{C}}}^{^*}\ \ \ ,\ \ \ i=1,2,3.  \color{white}\bigg]\ \ }  
\label{eq:HBB}  
\end{equation}   

This is crucial to our analysis since \eqref{eq:HBB} can be inverted to obtain the Higgs branch as a hyperk$\ddot{\text{a}}$ler quotient 

\begin{equation}   
\boxed{\ \ \ W_{_{G_{_{\mathbb{C}}}}}\ \equiv \ \mu^{^-1} /G^{^3}\color{white}\bigg]\ \ }.  
\label{eq:HBtrue}  
\end{equation}

However, for the case in which the identity is removed from the source category, \eqref{eq:HBtrue}  does not hold anymore precisely because of the lack of permutational symmetry arising in the quotient. In \cite{VP} we will be explaining how to define $W_{_{G_{_{\mathbb{C}}}}}$ and its dimensionality for the case involving categories without a duality structure.

\section*{Key points}    

The main points are the following:  

\begin{itemize}  

\item Categorical duality ensures the presence of an identity object.

\item S-duality requires reparametrisation-invariance.

\end{itemize}

\section{Moore-Tachikawa varieties beyond duality}   \label{sec:new1234}

In this concluding section, we piece together several considerations made throughout our treatment, ultimately showing how apparent shortcomings in mathematical descriptions might lead to interesting physical realisations.  

This section is structured as follows:

\begin{enumerate}

\item We show why the formalism of \cite{Moore:2006dw} needs to be generalised if reparametrisation-invariance of the Riemann surface falls short from being satisfied, and how this opens up to interesting generalisations of the proposal of \cite{Moore:2011ee} for 2D TFTs describing maximal dimensional Higgs branches of class ${\cal S}$ theories\footnote{The latter will be the core topic of section \ref{sec:new1234}.}.  In particular, we build the relation with the definition of the fiber functor and Drinfeld center, \eqref{eq:HB1}.

\item We conclude highlighting interesting features of the formalism of \cite{Moore:2006dw,Moore:2011ee} in absence of reparametrisation-invariance of the Riemann surface, connecting them to the emergence of intrinsically non-invertible symmetries separating different class ${\cal S}$ theories, \cite{Pasquarella:2023deo}.

\end{enumerate}

\subsection{Categorical structure without duality}   \label{sec:newcat}

As explained in the previous section, removing the identity element in Bo$_{_2}$ requires having to introduce at least two different functors $\eta_{_{G_{_{\mathbb{C}}}^{^{\prime}}}}, \eta_{_{G_{_{\mathbb{C}}}^{^{\prime\prime}}}}$, whose action on the circle and its bordism reads as follows 

\begin{equation}   
\eta_{_{G_{_{\mathbb{C}}}^{^{\prime}}}}\left(\ S^{^1}\ \right)\ \equiv\ G_{_{\mathbb{C}}}^{^{\prime}}\ \ \ , \ \ \ \eta_{_{G_{_{\mathbb{C}}}^{^{\prime\prime}}}}\left(\ S^{^1}\ \right)\ \equiv\ G_{_{\mathbb{C}}}^{^{\prime\prime}}   
\label{eq:morph0}  
\end{equation} 

\begin{equation}   
\eta_{_{G_{_{\mathbb{C}}}^{^{\prime}}}}\left(\ U\ \right)\ \equiv\ U_{_{G_{_{\mathbb{C}}}^{^{\prime}}}} \equiv \  G_{_{\mathbb{C}}}^{^{\prime}}\times S_{_n}\ \subset \ G_{_{\mathbb{C}}}^{^{\prime}}\times\mathfrak{g}_{_{\mathbf{C}}}^{^{\prime}}\ \simeq \ T^{^*}G_{_{\mathbf{C}}}^{^{\prime}},       
\label{eq:morph1}  
\end{equation}

\begin{equation}   
\eta_{_{G_{_{\mathbb{C}}}^{^{\prime\prime}}}}\left(\ U\ \right)\ \equiv\ U_{_{G_{_{\mathbb{C}}}^{^{\prime\prime}}}} \equiv \  G_{_{\mathbb{C}}}^{^{\prime\prime}}\times S_{_n}\ \subset \ G_{_{\mathbb{C}}}^{^{\prime\prime}}\times\mathfrak{g}_{_{\mathbf{C}}}^{^{\prime\prime}}\ \simeq \ T^{^*}G_{_{\mathbf{C}}}^{^{\prime\prime}},  
\label{eq:morph2}   
\end{equation}

\begin{equation}   
\eta_{_{G_{_{\mathbb{C}}}^{^{\prime}}}}, \left(\ V\ \right)\ \equiv\ V_{_{G_{_{\mathbb{C}}}^{^{\prime}}}} \overset{def.}{\equiv} \  T^{^*}G_{_{\mathbb{C}}}^{^{\prime}},      
\label{eq:id1}  
\end{equation}

\begin{equation}   
\eta_{_{G_{_{\mathbb{C}}}^{^{\prime\prime}}}}\left(\ V\ \right)\ \equiv\ V_{_{G_{_{\mathbb{C}}}^{^{\prime\prime}}}} \overset{def.}{\equiv} \  T^{^*}G_{_{\mathbb{C}}}^{^{\prime\prime}},  
\label{eq:id2}   
\end{equation} 
where we are assuming

\begin{equation}   
\mathfrak{g}_{_{\mathbf{C}}}^{^{\prime}}\ \cap\ \mathfrak{g}_{_{\mathbf{C}}}^{^{\prime\prime}}\ \neq\ \{ \emptyset \}\   \ , \ \ \text{and}\ \ \   \mathfrak{g}_{_{\mathbf{C}}}^{^{\prime}}\ \cup\ \mathfrak{g}_{_{\mathbf{C}}}^{^{\prime\prime}}\ \equiv\ \mathfrak{g}_{_{\mathbf{C}}}, 
\label{eq:fa}
\end{equation}

\begin{equation}   
\mathfrak{g}_{_{\mathbf{C}}}^{^{\prime}}\ \not\subset\ \mathfrak{g}_{_{\mathbf{C}}}^{^{\prime\prime}}\    \ , \ \ \text{and}\  \  \ \ \mathfrak{g}_{_{\mathbf{C}}}^{^{\prime\prime}}\ \not\subset\ \mathfrak{g}_{_{\mathbf{C}}}^{^{\prime}}.    
\label{eq:sa}
\end{equation} 

\eqref{eq:fa} and \eqref{eq:sa} imply that the two subalgebras involved, $\mathfrak{g}_{_{\mathbf{C}}}^{^{\prime}}, \mathfrak{g}_{_{\mathbf{C}}}^{^{\prime\prime}}$ are associated to different subgroups, $G_{_{\mathbb{C}}}^{^{\prime}}, G_{_{\mathbb{C}}}^{^{\prime\prime}} $, and that the identity elements differ, $T^{^*}G_{_{\mathbf{C}}}^{^{\prime}}\ \neq\ T^{^*}G_{_{\mathbf{C}}}^{^{\prime\prime}}$, even under reparametrisation of the Reimann surface. For each one of the 2D TFTs, $\eta_{_{G_{_{\mathbb{C}}}^{^{\prime}}}}, \eta_{_{G_{_{\mathbb{C}}}^{^{\prime\prime}}}}$ one could use the formalism of \cite{Moore:2006dw,Moore:2011ee}, describing two different class ${\cal S}$ theories, both descending from 6D ${\cal N}=(2,0)$ by dimensionally reducing on a Riemann surface without reparemetrisation invariance. However, given the assumption that $(G_{_{\mathbb{C}}}^{^{\prime}}, \mathfrak{g}_{_{\mathbf{C}}}^{^{\prime}}), (G_{_{\mathbb{C}}}^{^{\prime\prime}}, \mathfrak{g}_{_{\mathbf{C}}}^{^{\prime\prime}}) $ can be embedded in a unique $(G_{_{\mathbb{C}}}, \mathfrak{g}_{_{\mathbb{C}}})$, it is natural to ask what should the triple on the RHS of \eqref{eq:HB1} be for the resulting theory to be absolute?  

We know that the fiber functor and Drinfeld centers for a given absolute theory are defined as follows  

\begin{equation}{\cal F}\ :\ \mathfrak{Z}\left(\text{Bo}_{_2}\right)\ \rightarrow\ \text{Bo}_{_2}, 
\label{eq:F}
\end{equation} 
         
      \begin{equation}  
\mathfrak{Z}\left(\text{Bo}_{_2}\right)\equiv\text{End}_{_{\text{Bo}_{_2}}}\left(T^{^*}G_{_{\mathbf{C}}}\right)\equiv\text{Hom}_{_{\text{Bo}_{_2}}}\left(T^{^*}G_{_{\mathbf{C}}},T^{^*}G_{_{\mathbf{C}}}\right),    
\label{eq:Z}
\end{equation}  
both of which crucially rely upon the presence of an identity in the source and target of $\eta_{_{G_{_{\mathbb{C}}}}}$. Apparently, we run into a contradiction, since the identity element $T^{^*}G_{_{\mathbf{C}}}$ has been removed by assumption, thereby implying $\mathfrak{Z}\left(\text{Bo}_{_2}\right)$ cannot be defined in the ordinary way.   On the other hand, one could define the new identity as being a composite object defined in the following way

      \begin{equation}  
\boxed{\ \ \ T^{^*}\tilde G_{_{\mathbf{C}}}\ \overset{def.}{=}\ T^{^*}G_{_{\mathbf{C}}}^{^{\prime}}\ \otimes_{_{T^{^*}G_{_{\mathbb{C}}}}}\ T^{^*}G_{_{\mathbf{C}}}^{^{\prime\prime}}\color{white}\bigg]\ \ },    
\label{eq:newid}
\end{equation} 
and therefore our proposal for \eqref{eq:F} and \eqref{eq:Z} reads as follows

\begin{equation}
\boxed{\ \ \ {\cal F}\ :\ \mathfrak{Z}\left(\tilde{\text{Bo}_{_2}}\right)\ \rightarrow\ \tilde{\text{Bo}_{_2}}\color{white}\bigg]\ \ }, 
\label{eq:Fnew}
\end{equation} 
         
      \begin{equation}  
\boxed{\ \ \ \mathfrak{Z}\left(\tilde{\text{Bo}_{_2}}\right)\ \equiv\ \text{End}_{_{\tilde{\text{Bo}_{_2}}}}\left(T^{^*}G_{_{\mathbf{C}}}^{^{\prime}}\ \otimes_{_{T^{^*}G_{_{\mathbb{C}}}}}\ T^{^*}G_{_{\mathbf{C}}}^{^{\prime\prime}}\right)\color{white}\bigg]\ \ },    
\label{eq:Znew}
\end{equation}
with 

\begin{equation}  
\tilde{\text{Bo}_{_2}}\ \overset{def.}{=}\ \text{Bo}_{_2}\ /\ V.
\end{equation}  

In the following subsection, we will explain why \eqref{eq:Fnew} and \eqref{eq:Znew} are reasonable proposals for defining an absolute theory in absence of a categorical duality structure.

\subsection{Interesting shortcomings}

As explained in \cite{Pasquarella:2023deo}, the loss of reparametrisation invariance of the Riemann surface signals the presence of an intrinsically-non-invertible defect between different class ${\cal S}$ theories. We will now show that the setup described in section \ref{sec:newcat} is equivalent to that of \cite{Pasquarella:2023deo}.

The starting point in our argument is the conjecture \eqref{eq:HBB} and \eqref{eq:HBtrue}. In absence of categorical duality of, both, source and target, the 2D TFT associated to the maximal dimensional Higgs branch of \cite{Moore:2011ee} is no longer associated to a moment map that is equivalent for all the constituent $S^{^1}$s, thereby violating the conjecture made by \cite{Moore:2011ee}. Explicitly,

\begin{equation}   
\boxed{\ \ \ \ W_{_{G_{_{\mathbb{C}}}}}\ \neq\  \mu^{^-1} /G^{^3}\color{white}\bigg]\ \ } . 
\label{eq:HB}  
\end{equation}  

This is because the algebraic variety associated to the Higgs branch $W_{_{G_{_{\mathbb{C}}}}}$ is not a hyperk$\ddot{\text{a}}$ler quotient. In particular, it is associated to a non-primitive ideal. 

Expanding further on this topic, the crucial point is that, when giving up the duality propriety, there is no longer the identity element in the source category, $T^{^*}G_{_{\mathbf{C}}}$, but, rather, there is one identity for each underlying constituent, \eqref{eq:morph1} and \eqref{eq:morph2}.  

In order to determine how the resulting composite identity element should look like, we need to briefly recall what was outlined in section \ref{sec:od}. Given two different morphisms, and taking their composition

\begin{equation}  
T^{^*}G_{_{\mathbf{C}}}^{^{a^{^{\prime}}}}\ \circ\ T^{^*}G_{_{\mathbf{C}}}^{^a}\equiv\left(T^{^*}G_{_{\mathbf{C}}}^{^a}\ \times\ T^{^*}G_{_{\mathbf{C}}}^{^{a^{^{\prime}}}} \right)//G_{_{\mathbf{C}}},  
\label{eq:compths}   
\end{equation}   
we know that, under the duality assumption, the axiom \eqref{eq:composition} 
comes with two moment maps, \eqref{eq:etaGC1}, each one describing the embedding of the individual morphisms within the algebra of the mother theory. As explained in section \ref{sec:od}, one actually uses the mutual relation in between such moment maps to prove that $T^{^*}G_{_{\mathbf{C}}}$ behaves as the identity. As also claimed in the previos section, \eqref{eq:compths} is expresses the need to define a Drinfeld center for the composite system made up of two class $\cal S$ theories. In presence of reparametrisation invariance, such theories can be thought of as being separated by an invertible defect, from which \eqref{eq:compths} can be reduced to a statement of S-duality. 

On the other hand, in absence of reparametrisation-invariance, the resulting class ${\cal S}$ theories would, by definition, be separated by an intrinsically non-invertible defect, with the latter being responsible for the lack of reparametrisation invariance of the Riemann surface \cite{Moore:2006dw}.

Our main question is to find the Drinfeld center for a given Bo$_{_2}$       

\begin{equation}  
{\cal F}:\ \mathfrak{Z}(\text{Bo}_{_2})\ \rightarrow\ \text{Bo}_{_2} 
,
\label{eq:fiber}  
\end{equation}  
and we know that, to a given fiber functor, ${\cal F}$, there is an associated moment map 

\begin{equation}  
\mu:\ \mathfrak{G}\ \rightarrow\ {\cal A}.     
\end{equation}  
with $\mathfrak{G}$ being Bo$_{_2}$ in this case, and ${\cal A}$ the algebra of invertible topological defects that projects to the identity $T^{^*}G_{_{\mathbb{C}}}$ under complete gauging of the theory. This is equivalent to stating that $T^{^*}G_{_{\mathbb{C}}}$ is the identity element once having projected over ${\cal A}$

\begin{equation}  
{\cal A}\ //_{\mu}\ \mathfrak{G}\ \ \ \text{with}\ \ \ \mu:\ \mathfrak{G}\ \rightarrow\ {\cal A}  
\end{equation}   
choosing the definition of the identity in the following way 

\begin{equation}  
T^{^*}G_{_{\mathbb{C}}}\ \simeq\ \mathbb{1}_{_{G_{_{\mathbb{C}}}}}\ \equiv\ {\cal A}\ //_{\mu}\ \mathfrak{G}.    
\end{equation} 

For the purpose of our work, $\mathfrak{G}\ \overset{def.}{=}\ \text{Bo}_{_2}$, therefore    

\begin{equation}  
T^{^*}G_{_{\mathbb{C}}}\ \simeq\ \mathbb{1}_{_{G_{_{\mathbb{C}}}}}\ \equiv\ {\cal A}\ //_{\mu}\ \text{Bo}_{_2}
\end{equation}  
However, in the case of section \ref{sec:newcat}, there are two different moment maps involved, one for each choice of conformal structure on the Riemann surface, that are not mutually related by the moment map constraint following from the axiom \eqref{eq:composition}. We therefore need to find the moment map (and corresponding gauge group $\tilde G_{_{\mathbb{C}}}$)

\begin{equation}  
\tilde\mu:\ \tilde{\mathfrak{G}}\ \rightarrow\ \tilde{\cal A},        
\end{equation}
whose identity element

\begin{equation} 
\tilde{\cal A}\ \overset{def.}{=}\ {\cal A}_{_1}\ \otimes_{_{T^{^*}G_{_{\mathbb{C}}}}}{\cal A}_{_{2}}    
\end{equation}  
constitutes the identity of the composite theory. If $T^{^*}G_{_{\mathbb{C}}}$ is the identity that has been removed from a particular source category, it still exists, but is no longer the identity present in $\tilde{\text{Bo}}$$_{_2}$ of a given $\tilde\eta_{_{G_{_{\mathbb{C}}}}}$. From the RHS of \eqref{eq:compths}, the identity of $\tilde{\text{Bo}}_{_2}$ therefore reads

      \begin{equation}  
      T^{^*}\tilde G_{_{\mathbf{C}}}\ \overset{def.}{=}\ 
T^{^*}G_{_{\mathbf{C}}}^{^{\prime}}\ \otimes_{_{T^{^*}G_{_{\mathbb{C}}}}}\ T^{^*}G_{_{\mathbf{C}}}^{^{\prime\prime}},    
\label{eq:newidnow}
\end{equation} 
such that its Drinfeld center can be determined.

Defining $\tilde\mu:\ \tilde{\text{Bo}}_{_2}\rightarrow\ \tilde{\cal A}$ as the moment map associated to the composite theory, the corresponding fiber functor can be explicitly rewritten as follows  

\begin{equation}  
\boxed{\ \ \ {\cal F}:\ \mathfrak{Z}\left(\tilde\eta_{_{\tilde G_{_{\mathbb{C}}}}}^{^{-1}}\left(T^{^*}\tilde G_{_{\mathbb{C}}}\right)\right)\ \rightarrow\ \tilde\eta_{_{\tilde G_{_{\mathbb{C}}}}}^{^{-1}}\left(T^{^*}\tilde G_{_{\mathbb{C}}}\right)\color{white}\bigg]\ \ }.   
\label{eq:fiber}  
\end{equation}  
ultimately enabling us to reformulate the problem of finding the Drinfeld center to that of identifying ${\cal F}$ for a given $\mu$.

\section*{Key points}    

The main points are the following:  

\begin{itemize}  

\item Lack of reparametrisation-invariance of bordism operators signals the presence of intrinsic non-invertible defects separating different class ${\cal S}$ theories.

\item Defining the Drinfeld center for a system of composite class ${\cal S}$ theories separated by non-invertible defects constitutes a nontrivial generalisation of an S-duality statement.

\end{itemize}

\section{Conclusions and Outlook}

The present article is the first of a series of works by the same author providing mathematical support of the claims made in \cite{Pasquarella:2023deo, Pasquarella:2023exd}. 

At first, we briefly overviewed cochain level theories as the most important generalisation of the open and closed TFT construction, emphasising its relation to the SymTFT construction leading to absolute theories. Mostly relying upon \cite{Moore:2006dw}, we highlighted the importance of the construction of cobordism operators, emphasising their dependence on the conformal structure of the Riemann surface. In section \ref{sec:BMTCs} we then turned to the discussion of a particular 2D TFT valued in a symmetric monoidal category, namely the maximal dimension Higgs branch of class ${\cal S}$ theories. After briefly reviewing the properties outlined in \cite{Moore:2011ee}, we propose their generalisation for the case in which the target category of the $\eta_{_{G_{_{\mathbb{C}}}}}$ functor is a hyperk$\ddot{\text{a}}$hler quotient. We concluded outlining the possible extension of this treatment towards a mathematical formulation of magnetic quivers within the context of Coulomb branches of 3D ${\cal N}=4$ quiver gauge theories which will be addressed in an upcoming work by the same author.

\section*{Acknowledgements}    

I wish to thank Hiraku Nakajima and Constantin Teleman for insightful discussions on related topics. My work is partly supported by an STFC studentship through DAMTP.

\appendix

\end{document}